\documentclass{PoS}
\usepackage{amsfonts}
\usepackage{epsfig}
\usepackage{float}
\usepackage{latexsym}
\usepackage{graphicx}
\usepackage{amssymb}
\usepackage{amsmath}
\usepackage{color}
\usepackage{footnote}

\title{The AdS/CFT relation, quasi-normal modes and applications}

\ShortTitle{AdS/CFT applications}

\author{\speaker{E. Abdalla,$^a$} C. E. Pellicer$^a$ and Jeferson de Oliveira$^b$\\
        \llap{$^a$} Instituto de F\'\i sica, Universidade de S\~ao Paulo\\
        CEP 05315-970, S\~ao Paulo, Brazil\\
        \llap{$^b$}Instituto de F\'\i sica, Universidade Federal do Mato Grosso,\\
        CEP 78060-900, Cuiab\'a, Brazil
        E-mail: \email{eabdalla@usp.br},\email{pellicer@usp.br},\email{jeferson@fisica.ufmt.br}}

\abstract{
Here we present a fast review of some of developments and
new results concerning applications of gravity in the context of
the AdS/CFT correspondence, in brane world perturbations as well as
in holographic superconductors. 
We also discuss the structure of
the phase transitions in a more general
set up defined by the quasi-normal oscillations in the bulk, which
signalize a very complex structure of the
phase transition at the border.
}

\FullConference{7th Conference Mathematical Methods in Physics - Londrina 2012,\\
		16 to 20 April 2012\\
		Rio de Janeiro, Brazil}

\begin{document}




\section*{Introduction}
Applications of gravity outside the strict realm of Einstein gravity
itself is a growing field of research, especially
after the discovery of the Anti de Sitter/ Conformal Field Theory
(AdS/CFT) relation \cite{maldacena1,witten}
and its implications for condensed matter systems \cite{1H}. In
particular, methods initially used to
treat question which concern mainly Einstein gravity, such as quasi-normal 
perturbations for the problem
of stability of black holes, or the behaviour of electromagnetic
fields in their vicinity, found applications
in the AdS/CFT context to find, respectively, phase transitions
\cite{gubser,phasetrans} or superconductivity
\cite{1H,3Hs}. With the growing of the complexity of the
problems, it became natural to consider
the same methods for any gravitational model in asymptotically AdS
space and the perturbations thereof, as
a means to check further properties of further conformally invariant
theories at the border of the AdS space.

Indeed, after the holographic principle was suggested by 't Hooft
\cite{thooft} and Susskind \cite{susskind},
a large amount of literature emerged, culminating with a more precise
though still conjectured formulation,
where a theory of gravity in $d$ dimensions is equivalent to a
conformally invariant theory in a $d-1$
dimensional space-time defined at the AdS infinity.

Later, the fact that a conformally invariant field theory is a natural
set up for describing critical phenomena
was used to define new models of the latter by means of the higher
dimensional gravity counterpart. Thus,
it has been proved, using the AdS/CFT conjecture and the so called
dictionary relating fields defined at
the infinity AdS border to the fields in the CFT counterpart, that a
Reissner-Nordstr\"om black hole surrounded
by electromagnetic fields and a scalar leads to a model of
superconductivity at the border, where the scalar
plays the role of an order parameter and one can compute the current
and electromagnetic field at the CFT theory
by means of the dictionary thus computing the conductivity. The
amazing part of such a story is the fact that
one need to know just classical gravity in the so called bulk to
arrive at such conclusions about an eminently
quantum theory at the border.


\section{AdS Setting}

A fundamental problem in General Relativity is to find a metric $g_{\mu\nu}$ that is a solution to the Einstein's equations
for a given energy-momentum tensor $T_{\mu\nu}$. The Minkowsky metric $g_{\mu\nu} = \eta_{\mu\nu} = \operatorname{diag}(-1,1,1,1)$
is a trivial solution for $T_{\mu\nu}=0$. The line element $ds^2 = \eta_{\mu\nu} dx^\mu dx^\nu$ is invariant under transformations from
the group $\operatorname{SO}(1,3)$, or $\operatorname{SO}(1,d-1)$ in $d$ dimensions. 

Taking the Einstein's equations in the vacuum and adding a cosmological term $8 \pi T_{\mu\nu} = - \Lambda g_{\mu\nu}$,
where $\Lambda$ is the cosmological constant, and contracting, we see that the metric $g_{\mu\nu}$ describes a $d$-dimensional 
space-time with constant curvature $\left( 1 - \frac{d}{2} \right) R = - d \Lambda $.
If $R>0$, the space-time is called de Sitter space (dS), and if $R<0$, Anti de Sitter space (AdS).  

The solution for the Einstein's equations is
\begin{equation}
ds^2 = - \left( 1 + \frac{r^2}{L^2} \right) dt^2 + \frac{1}{\left( 1 + \frac{r^2}{L^2} \right)} + \frac{r^2}{L^2} d\Omega^2 \ ,
\end{equation}
with
$L^2 = - (d-1)(d-2)/(2\Lambda)$
and $d\Omega^2$ is a unitary $2$-sphere metric.

This $d$-dimensional space can be defined as a embedding in a $d+1$-dimensional space with line element
\begin{equation}
ds^2 = -dt^2 + \sum_{i=1}^{d-1} dx_i^2 - dx_d^2
\end{equation}
constrained by
\begin{equation}
-t^2 + \sum_{i=1}^{d-1} x_i^2 - x_d^2 = -L^2 \ .
\end{equation}
This constraint is invariant by transformation from the conformal groups $\operatorname{SO}(2,d-1)$, so this is the symmetry group
of the AdS space. For further information, see \cite{nastase}.

The metric of the AdS space can be expressed in Poincar\'e coordinates, 
\begin{equation}\label{poincare}
ds^2 = \frac{L^2}{z^2} \left( dz^2 + \eta_{\mu\nu} dx^\mu dx^\nu \right) \ ,
\end{equation}
where $z \in (0,\infty)$ is an inverse radial coordinate. A free scalar field $\psi(z,x^\mu)$ is a solution to the Klein-Gordon
equation
\begin{equation}
\square \psi - m^2 \psi = 0 \ .
\end{equation}
Since the metric~(\ref{poincare}) does not depend on the coordinates $x^\mu$, we can decompose the scalar field as
$\psi(z,x^\mu) = e^{ik_\mu x^\mu} \psi_k(z)$
and write the Klein-Gordon equation as
\begin{equation}\label{eq_kg}
\frac{1}{L^2} \left[ z^2 k^2 - z^{d+1} \partial_z \left( z^{-d+1} \partial_z \right) + m^2 L^2 \right] \psi_k(z) = 0 \ ,
\end{equation}
as seen on \cite{mcgreevy}.

As $z\to 0$, we can neglet the term $z^2 k^2$ in eq.~\eqref{eq_kg} and $\psi_k$ has power law solutions $\psi_k = z^\Delta$ with
$\Delta(\Delta-d) = m^2L^2$. So the asymptotic solution of $\psi(z,x^\mu)$ can be written as
\begin{equation}\label{asymp_z}
\psi(z,x^\mu) \simeq \alpha(x^\mu) z^{\Delta_-} + \beta(x^\mu) z^{\Delta_+}
\end{equation}
with
\begin{equation}\label{delta_pm}
\Delta_\pm = \frac{d}{2} \pm \sqrt{\left( \frac{d}{2} \right)^2 + m^2 L^2 } \ .
\end{equation}

\section{Perturbations around black hole solutions}

Let $g^{(0)}_{\mu\nu}$ be a known black hole solution. To this metric we add a small perturbation $h_{\mu\nu}$, whose components must 
be small compared to $g^{(0)}_{\mu\nu}$, so the metric is now
\begin{equation}
g_{\mu\nu} = g^{(0)}_{\mu\nu} + h_{\mu\nu} \ .
\end{equation}
To calculate $h_{\mu\nu}$, we write the Einstein's equations for $g_{\mu\nu}$ in terms of $g^{(0)}_{\mu\nu}$ and $h_{\mu\nu}$, isolate the terms 
depending only on $g^{(0)}_{\mu\nu}$, which must cancel since $g^{(0)}_{\mu\nu}$ is already a solution to Einstein's equations, and neglet the 
terms proportional to $h^2$ and higher orders to derive linear differential equations for $h_{\mu\nu}$. 
For good reviews, see \cite{kokkotas,nollert}.

In a spherically symmetric space-time, it is possible to decouple the angular variables and define a scalar function $\psi(r,t)$ as a 
function of some component of $h_{\mu\nu}$, so we can solve the differential equations for each component of $h_{\mu\nu}$ solving a 
partial differential equation for a scalar field $\psi(r,t)$. In a convenient coordinate choice $r_\star (r)$, this equation is
\begin{equation}
-\frac{\partial^2 \psi}{\partial t^2} + \frac{\partial^2 \psi}{\partial r_\star^2} - V_{\text{eff}} (r) \psi = 0 \ ,
\end{equation}
where $V_{\text{eff}}(r)$ is the effective potential depending only on the background metric $g^{(0)}_{\mu\nu}$.

The behaviour of $\psi$ can be expressed as $\psi(r,t) = e^{-i\omega t}\psi(r)$. These functions are called quasi-normal modes because they
may lack properties of normal modes, such as completeness. The quasi-normal frequency $\omega$ is complex, and the quasi-normal mode is stable 
if $\Im(\omega)<0$ and unstable if $\Im(\omega)>0$. It is possible to prove that is the effective potencial $V_{\text{eff}}(r)$ is positive
for all $r$, the quasi-normal modes are stable\cite{wald,2Hs}. 
For unstable mode to appear, it is necessary that the potential has pits with bound states of 
negative energy \cite{franceses}. 

\section{Black $p$-brane solutions}
The black $p$-branes are a class of extended  solutions of the bosonic sector of  
IIB/IIA Supegravity \cite{gibbons}. The near horizon limit of such geometries is 
given by the product between the $(p+2)$-dimensional AdS space-time 
and an $(d-p-2)$-dimensional sphere S:  $AdS_{p+2}\times S^{d-p-2}$, where $d$ is 
the number of space-like dimensions in the bulk and $p$ is the number of $p$-brane 
extended dimensions. Also, the spacelike infinity is the $d$-dimensional Minkowski 
space-time, showing that the black $p$-branes solutions interpolate the AdS spacetime 
and the flat geometry. This connection, at least in ten dimensions,  was very important 
for the first version of AdS/CFT conjecture as presented by Maldacena \cite{maldacena1}.

As shown in detail in our previous work \cite{pbranas}, the black $p$-branes solution 
can be regarded as black holes with extended event horizons. So, we can apply the 
well-known perturbations methods of black holes in general relativity  to the present 
case, in order to obtain some information about its structure and stability.

The ten dimensional Supergravity solution describing the so-called black p-branes 
is given by metric
\begin{equation}\label{brane}
ds^{2}=-A(x)dt^{2}+B(x)\left[dr^{2}+r^{2}d\Omega^{2}_{p-1}\right]+C(x)dx^{2}
+x^{2}D(x)d\Omega^{2}_{8-p},
\end{equation}
where $A(x)=(1-(a/x)^{7-p})(1-(b/x)^{7-p})^{-1/2}$, 
$C(x)=(1-(b/x)^{7-p})^{\alpha_1}(1-(a/x)^{7-p})^{-1}$, $B(x)=\sqrt{1-(b/x)^{7-p}}$, 
$D(x)=(1-(b/x)^{7-p})^{\alpha_2}$, with $\alpha_1=-\frac{1}{2}-\frac{(5-p)}{(7-p)}$ 
and $\alpha_2=\frac{1}{2}-\frac{(5-p)}{(7-p)}$. The maximal extension of above 
metric describes a black hole geometry, with an event horizon located at $x=a$. 
In the case, $p\neq3$, a physical singularity in the curvature is present at $x=b$. 
Otherwise,  we have in addition to the event horizon at $x=a$, and inner horizon at 
$x=b$ and the singularity at $x=0$.

We consider as linear perturbation the massless scalar field $\Psi(x^{A})$ evolving 
in the geometry (\ref{brane}). In order to study the stability of black $p$-branes 
we obtain the scalar quasi-normal spectrum. In the work \cite{pbranas} we have applied 
two different numerical approaches, the WKB technique and the \emph{time-domain} 
approach. Here we will discuss only the \emph{time-domain} numerical results, since 
we have a good agreement between the methods. 

The Klein-Gordon equation for $\Psi(x^{A})$ in the metric (\ref{brane}) reduces to
\begin{equation}\label{kg1}
\frac{d^{2}}{dr_{*}^{2}}\Psi_{L}(x)+\left[k^{2}-V(x)\right]\Psi_{L}(x)=0,
\end{equation}
where $k^{2}=\omega^{2}-\beta^{2}$, and $V(x)$ is the effective potential, which is a 
well-behaved function which goes to zero at spatial infinity and has a peak near the 
event horizon.

 Following the gauge invariant formalism \cite{kodama}, we can make an important 
observation: the equation governing  the evolution of tensorial perturbations in such 
formalism is the Klein-Gordon equation. So, the quasi-normal spectrum are the same and 
the tensorial stability/instability can be achieved through the ordinary scalar perturbation.

\begin{figure}[htbp]
\begin{center}
\includegraphics[height=4.5cm, width=15.0cm]{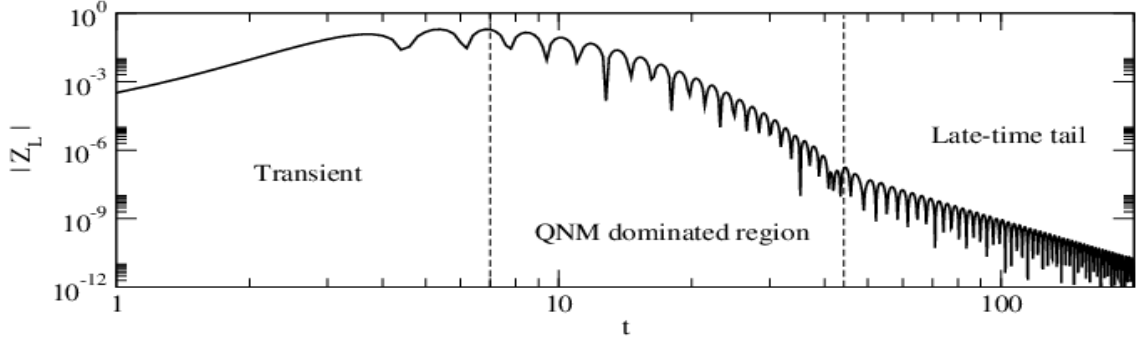}
\caption{{\bf Log-log graph of the absolute value of $Z_{L}(t,x_{fixed})$. The quasi-normal 
and tail phases are indicated. The $p$-brane parameters are $p=0$, $a=2$, $b=0.5$, $L=1$ 
and $\beta=1$}}
\label{fig1}
\end{center}
\end{figure}

In figure~\ref{fig1} we observed the usual picture in perturbative dynamics. After the 
initial transient regime, the quasi-normal mode phase follows as a late-time tail. Such an 
ending phase is strongly dependent on the value of the \emph{geometric mass} $\beta$.

As a conclusion, we can say that the black $p$-branes are stable against scalar and tensorial 
perturbations. Also, the quasi-normal spectrum in such a complex geometry is amazingly simple. 
Considering $\beta\neq 0$, later related to glueball mass in the AdS/CFT framework, the 
frequencies display an almost scaling behaviour. 

\section{The AdS/CFT Correspondence}

The $D3$-brane metric is
\begin{equation} 
ds^2 = \left( 1 + \frac{L^4}{r^4} \right)^{-\frac{1}{2}}
\left( - dt^2 + \sum_{i=1}^3 dx_i^2 \right) 
+ \left( 1 + \frac{L^4}{r^4} \right)^{\frac{1}{2}}
\left( dr^2 + r^2 d\Omega_5^2 \right) \ ,
\end{equation}
where $d\Omega_5^2$ is the $5$-sphere metric and $L^4 = 4 \pi g_s N^2$.
In an appropriate limit, this metric can be written as
\begin{equation}
ds^2 = \frac{r^2}{L^2} \left( -dt^2 + \sum_{i=1}^3 dx_i^2 \right)
+ \frac{L^2}{r^2} dr^2 + d\Omega_5^2 \ .
\end{equation}
With the coordinate change $r = L^2/u$, the metric takes the form
\begin{equation}
ds^2 = \frac{L^2}{u^2} \left( -dt^2 + \sum_{i=1}^3 dx_i^2 + du^2 \right) + L^2 d\Omega_5^2 \ ,
\end{equation}
which is the product of a $5$-dimensional AdS space with a $5$-sphere.

The conjecture stated by Maldacena says that a type IIB superstring theory
on AdS$_5\times$S$^5$ with string coupling $g_s$ and a $4$-dimensional
$N=4$ Super-Yang-Mills (SYM) theory with gauge group $SU(N)$ and SYM coupling $g_{YM}$
are equivalent, with $g_s=g_{YM}^2$ and $L^4 = 4\pi g_s N^2$.
Consider an operator $O$ in $N=4$ SYM, as seen in \cite{mcgreevy}. 
A generating funcional $Z[J]$ can be obtained perturbing
the lagrangian
$\mathcal{L}(x) \to \mathcal{L}(x) + J(x) O(x)$
so
\begin{equation}
Z[J] = \left< e^{-\int d^4 x J(x) O(x)} \right>_{\text{CFT}} \ .
\end{equation}
A deformation in the operator $O$ changes the value of $g_{YM}$. Since they are related, $g_s$ must also
change, which is equivalent to change the boundary value of the dilaton.

Let $\psi(z,x^\mu)$ be the dilaton, defined on the metric~(\ref{poincare}). 
The boundary value is written as $\psi(z=0,x^\mu) = \psi_0(x^\mu) = J(x)$.
According to the AdS/CFT correspondence \cite{agmoo},
\begin{equation}\label{gpkw}
\left< e^{-\int d^4 x J(x) O(x)} \right> = Z_{\text{string}} \left[ \psi(z,x^\mu) = \psi_0(x^\mu) \right] \ .
\end{equation}
$Z_{\text{string}}$ is the partition function on the AdS space where the field $\psi$ has the value 
$\psi_0$ as boundary condition.

Returning to the free scalar fields on AdS, whose asymptotic solution if given by
eq.~\eqref{asymp_z},
with $\Delta_\pm$ given by eq.~\eqref{delta_pm}, 
we choose one of the independent solution, that it, $\psi(z,x^\mu) = \alpha(x^\mu) z^{\Delta_-}$ and 
$\beta(x^\mu) = 0$ or $\psi(z,x^\mu) = \beta(x^\mu) z^{\Delta_+}$ and $\alpha(x^\mu)=0$.
Let $\epsilon$ be a coordinate value close to the boundary $z=0$. Then, 
$\psi(\epsilon,x^\mu) = \alpha(x) \epsilon^{\Delta_-}$ ( or
$\psi(\epsilon,x^\mu) = \beta(x) \epsilon^{\Delta_+}$). $\epsilon$ has dimension $[\text{length}]^{-1}$ and 
$\psi$ is dimensionless. Therefore, $\alpha(x^\mu)$ has mass dimension $\Delta_-$ (or  
$\beta(x^\mu)$ has mass dimension $\Delta_+$). Since $J(x)=\alpha(x^\mu)$ (or $J(x)=\beta(x^\mu)$),
eq.~\eqref{gpkw} implies that the operator $O$ has dimension $\Delta_-$ (or $\Delta_+$).

\section{Holographic Superconductors}

According to the AdS/CFT correspondence, a gravitational theory in the bulk is related to a field theory on the boundary. A space-time with a 
black hole is related to a thermal field theory, whose temperature is equal to the Hawking temperature of the black hole. Charged scalar fields 
in the bulk are related to condensates in the thermal field theory, so these scalar fields must behave like a order parameter if the thermal
field theory describes a superconductivity.

Consider the lagrangian from \cite{3Hs}
\begin{equation}\label{lag_3H}
\mathcal{L} = R + \frac{6}{L^2} - \frac{1}{4}F_{\mu\nu}^2 - |\nabla_\mu \psi - i q A_\mu \psi |^2 - m^2 |\psi|^2 \ .
\end{equation}
If we rescale $\psi \to \psi/q$ and $A_\mu \to A_\mu/q $, then the matter action has a $1/q^2$ factor, so a large $q$ supresses the
backreaction in the metric.

Considering the planar neutral black hole
\begin{equation}
ds^2 = -f(r) dt^2 + f^{-1}(r) dr^2 + r^2 (dx^2 + dy^2) \ ,
\end{equation}
where
\begin{equation}
f(r) = \frac{r^2}{L^2} - \frac{M}{r} \qquad \text{and} \qquad
T = \frac{3 r_h}{4\pi L^2} = \frac{3}{4\pi} \frac{M^{\frac{1}{3}}}{L^{\frac{4}{3}}} \ ,
\end{equation}
we assume that the fields depend only on the radial coordinate $A_\mu dx^\mu = \Phi(r) dt$, $\psi = \psi(r)$.
Then the field equations for $m^2L^2=-2$ become
\begin{eqnarray} \label{eq_psi_3h}
& \psi^{\prime\prime} + \left( \frac{f^\prime}{f} + \frac{2}{r} \right) \psi^\prime + \frac{\Phi^2}{f^2}\psi + \frac{2}{L^2 f} \psi = 0 \ ,\\
& \Phi^{\prime\prime} + \frac{2}{r}\Phi^\prime - \frac{2\psi^2}{f} \Phi = 0 \ . \label{eq_phi_3h}
\end{eqnarray}

At the event horizon $r=r_h$, the equations of motion are regular if $\phi(r_h)=0$ and $f^\prime(r_h) \psi^\prime(r_h) + \frac{2}{L^2} \psi(r_h) = 0$.
So, there is a two parameter family of solutions with regular horizons. Asymptotically,
\begin{eqnarray}\label{eq_asym_psi}
\psi &=& \frac{\psi^{(1)}}{r} + \frac{\psi^{(2)}}{r^2} + \cdots \ , \\
\Phi &=& -\mu - \frac{\rho}{r} + \cdots \ . \label{eq_asym_phi}
\end{eqnarray}

To understand the asymptotic behaviour of $\Phi(r)$, consider a charged black hole perturbation.
The time component of the electromagnetic potential behaves as a power law decay next to the AdS boundary and does
not correspond to a field in the dual theory, but fixes the charge density $\rho$ of a state. This component must be
zero at the event horizon, so a constant must be added, which can be understood in the dual field theory
as a sum of a chemical potential $\mu$.

For $\psi$, either falloff if normalizable. We need to impose the condition that either $\psi^{(1)}$ or $\psi^{(2)}$ vanishes. To do so, 
for each pair of boundary conditions $(\psi(r_h),\Phi(r_h))$ we numerically solve eqs.~\eqref{eq_psi_3h} and~\eqref{eq_phi_3h} 
using fourth order Runge-Kutta method and
for large $r$, fit the behaviour os $\psi(r)$ and $\Phi(r)$ as eqs.~\eqref{eq_asym_psi} and~\eqref{eq_asym_phi}, which are linear on 
the parameters. This procedure gives us a map $(\psi(r_h),\Phi(r_h))\to (\psi^{(1)},\psi^{(2)},\mu,\rho)$. Using the shooting method, we 
find the pair of boundary conditions for which $\psi^{(1)}$ or $\psi^{(2)}$ vanishes. When $\psi^{(2)}=0$, we plot $<O_1> = \sqrt{2} \psi^{(1)}$
against $T$ and when $\psi^{(1)}=0$, $<O_2> = \sqrt{2} \psi^{(2)}$ against $T$.

At this point, is important to mention that  equations of motion for the charged scalar field $\psi$ 
and electric potential $\Phi$ are invariant under two groups of scaling symmetries. The first one
($r\rightarrow\lambda r$, $t\rightarrow \lambda t$, $L\rightarrow\lambda L$, $q\rightarrow \lambda^{-1} q$, $m\rightarrow\lambda^{-1}m$),
allow us to set the value of AdS radius $L=1$. The second one
($r\rightarrow\lambda r$, $t\rightarrow \lambda^{-1} t$, $\Phi\rightarrow\lambda^{-1}\Phi$),
we can take $r_{h}=1$, which implies that the Hawking temperature is $T=3/4\pi$. Furthermore, the  
two groups of symmetry imply 
$T\rightarrow\lambda^{2}T$, and $\rho\rightarrow \lambda^{2}\rho$.
From this, we see that 
$T\propto\rho^{1/2}$.
Therefore, we can do calculations with $r_{h}=1$ and evaluate $T/\rho^{1/2}$ to eliminate the 
scaling symmetries that we used to set $r_{h}=1$.

Several curves appear, and they can be labelled by the number of times that $\psi(r)$ changes signal. 
Taking only cases in which $\psi(r)$ does not change signal, 
the figures in \cite{3Hs} are obtained.

$\psi(r)$ behaves like a order parameter, but this behaviour alone is not enough to see if
the phase transition is that of a superconductor or a superfluid. We need then to obtain
the conductivity $\sigma(\omega)$, where $\omega$ is the frequency. Perturbing the
spatial component of the electromagnetic potential $A_x$ and assuming
$A_x(r,t) = e^{-i \omega t} A_x(r)$, we get
\begin{equation}
A_x^{\prime\prime} + \frac{f^\prime}{f} A_x^\prime + \left( \frac{\omega^2}{f^2} 
- \frac{2\psi^2}{f} \right) A_x = 0 \ .
\end{equation}
As $r \to r_h$, $A_x \to f^{-\frac{i\omega}{3 r_h}}$. As $r\to\infty$,
\begin{equation}
A_x \simeq A_x^{(0)} + \frac{A_x^{(1)}}{r} + \cdots \ .
\end{equation}

According to \cite{1H}, the AdS/CFT dictionary says that the electric field in the CFT
is equal to the electric field in the bulk as $r$ tends to infinity ($E_x = - \dot{A}_x^{(0)}$),
and the first subleading term is dual to the induced current ($<J_x> = A_x^{(1)}$).
Applying Ohm's law,
\begin{equation}
\sigma(\omega) = \frac{<J_x>}{E_x} = - \frac{i}{\omega} \frac{A_x^{(1)}}{A_x^{(0)}} \ ,
\end{equation}
whose behaviour is presented in \cite{3Hs}.
For temperatures higher than the critical temperature, the real part of the conductivity 
is constant. For lower temperatures, there is a energy gap, as expected by BCS theory.

\section{Phase transitions in Reissner-Nordstr\"om holographic superconductors}

We consider the bulk action of a massive charged scalar field $\psi$ interacting with an abelian field $A_\mu$ generated by a charged
AdS black hole, as in \cite{gubser}.
We keep the lagrangian~\eqref{lag_3H} and now
the background space-time is a spherically symmetric Reissner-Nordstr\"om AdS black hole
\begin{equation}
ds^2 = -f(r) dt^2 + \frac{1}{f(r)} dr^2 + r^2 d\Omega^2
\end{equation}
with 
\begin{equation}
f(r) = 1 - \frac{2M}{r} + \frac{Q^2}{4r^2} + \frac{r^2}{L^2} \ .
\end{equation}
The abelian field is now fixed, $A_\mu dx^\mu = \Phi(r) dt$ with
\begin{equation}
\Phi(r) = \frac{Q}{r} - \frac{Q}{r_+} \ .
\end{equation}
Assuming that $\psi$ depends only on $r$, the equation of motion is
\begin{equation} \label{eommme}
\frac{1}{\sqrt{-g}} \partial_r \left( \sqrt{-g} g^{rr} \partial_r \psi \right) - \left( m^2 - \frac{q^2\Phi^2(r)}{f(r)} \right) \psi = 0 \ .
\end{equation}

Choosing coordinates in which $r_+=Q=1$, the Hawking temperature is given by
\begin{equation}\label{hawkingtemp}
T = \frac{3}{4\pi} \left( \frac{1}{4} + \frac{1}{L^2} \right)
\end{equation}
and eq.~\eqref{eommme} depends only on the parameters $m$, $q$ and $L$.

\begin{figure}[htbp]
\includegraphics[height=7.0cm, width=4.5cm,angle=-90]{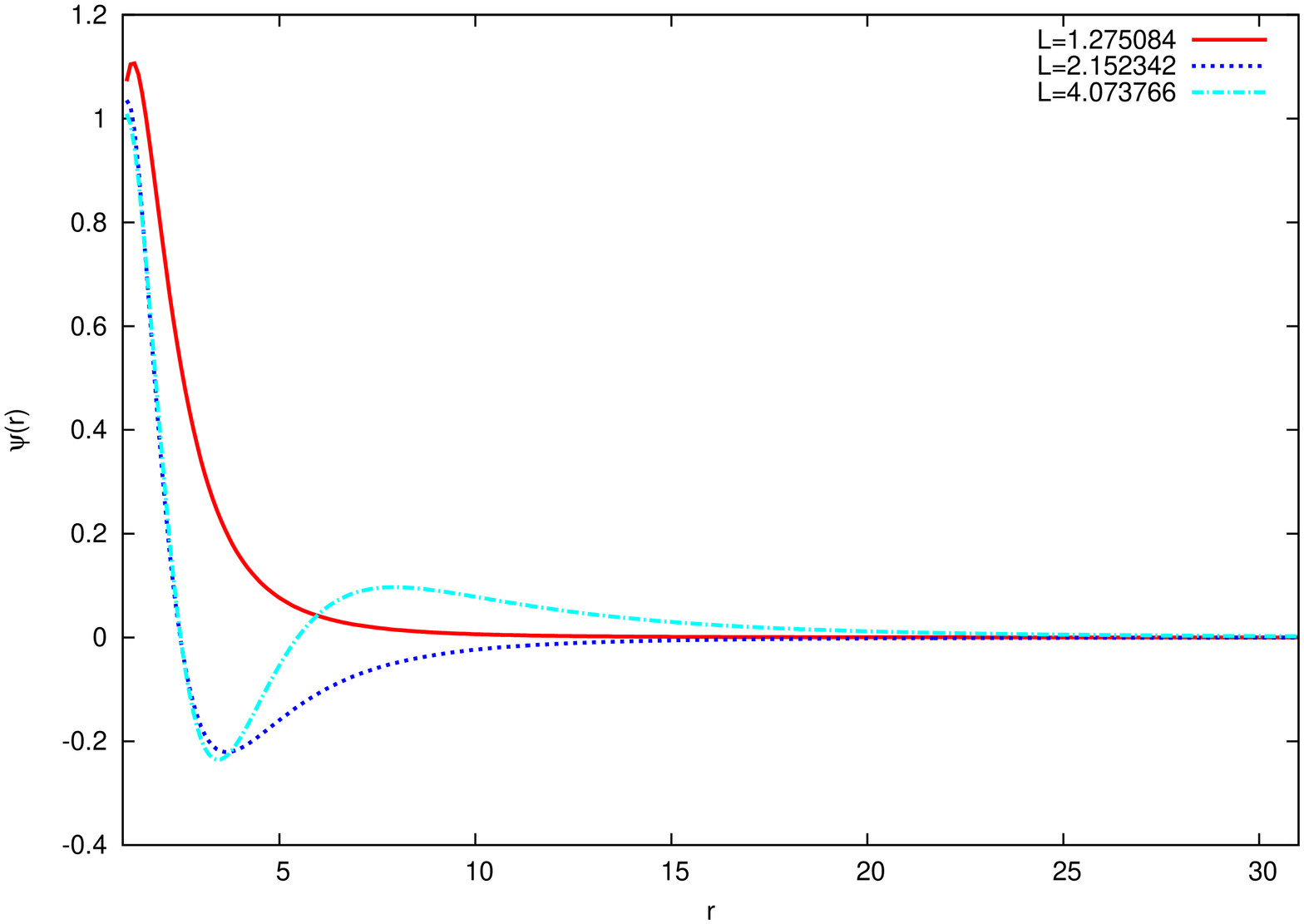}
\includegraphics[height=7.0cm, width=4.5cm,angle=-90]{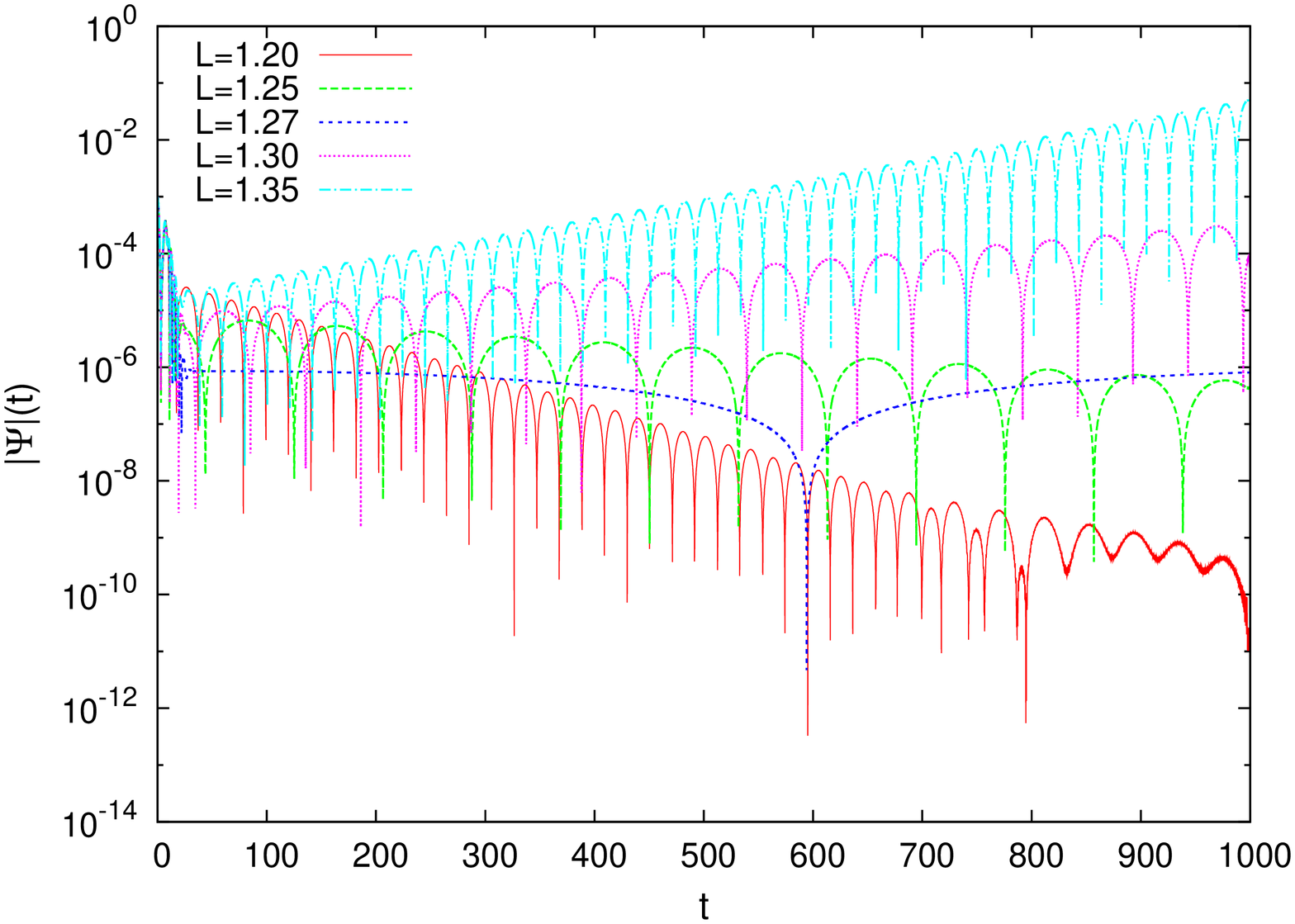}
\caption{{\bf Marginally stable modes (left) and quasi-normal modes near the transition (right)}}
\label{phasetrans1}
\end{figure}

We can set the boundary condition $\psi(r=1)=1$, since the equation of motion is linear. 
Developing a series solution around the horizon, we find $\psi^\prime(r=1) = \frac{4m^2L^2}{3L^2+12}$.
With these conditions, we can numerically solve eq.~\eqref{eommme} using fourth order Runge-Kutta method
and search parameters for which
$\psi \to 0$ as $r\to \infty$.  Fixing $m^2L^2=4$ and $qL=10$, we need to vary only the parameter $L$,
making the search straightforward. There are only three values of $L$ for which $\psi$ tends to zero, shown
in figure~\ref{phasetrans1}.
These modes can be labelled by the number of times $\psi$ changes signal and are called marginally stable modes.


Keeping the same fixed abelian field, but letting now $\psi$ depend on time as well as the radial coordinate, we
have the following equation of motion,
\begin{equation}
\square \psi - 2 i q A_\mu g^{\mu\nu} \partial_\nu \psi - q^2 A_\mu A^\mu \psi - m^2 \psi \ . 
\end{equation}
With the Ansatz $A_\mu dx^\mu=\Phi(r) dt$ and the tortoise coordinate $r_\star(r) = \int \frac{dr}{f(r)}$, we
write the equation por $\psi(r,t)$ as
\begin{equation}\label{eommqn}
- \frac{\partial^2 \psi}{\partial t^2} + \frac{\partial^2 \psi}{\partial r_\star^2} 
+ 2 i q \Phi(r) \frac{\partial \psi}{\partial t} - V_{\text{eff}}(r) \psi = 0
\end{equation}
with
\begin{equation}
V_{\text{eff}} (r) = f(r) \left( \frac{f^\prime(r)}{r} + m^2 \right) - q^2 \Phi^2(r) \ ,
\end{equation}
which is the equation for a charged scalar perturbation of the Reissner-Nordstr\"om-AdS black hole 
with an aditional imaginary term.

According to the AdS/CFT correspondence, a black hole is related to a thermal field theory, perturbations
of the black hole are perturbations of the thermal theory and the imaginary part of the quasi-normal frequencies
is related to the timescale of the return to thermal equilibrium \cite{2Hs}. 
For a condensate not to decay in the dual field theory, there must be an unstable quasi-normal mode in the bulk and the
phase transition must happen at the same temperature that the quasi-normal modes change stability.

Fitting the behaviour of the quasi-normal modes as $\psi(r,t) = e^{-i\omega t}\psi(r)$ and, if the frequency is zero,
as the imaginary part must be when the mode changes stability, it is easy to see that the
transition must happen at a marginally stable mode. At this point the meaning of more than 
one marginally stable mode is not clear.

We must now solve eq.~\eqref{eommqn} numerically. With a gaussian distribution as initial condition
and Dirichlet condition at $r_\star=0$, we derive the evolution of $\psi(r,t)$ using the finite difference
method. As in the case of marginally stable modes, we choose coordinates in which $r_+=Q=1$ and fix
$m^2L^2=4$ and $qL=10$ so the system depends only on one parameter $L$, which has a biunivocal relation with the Hawking 
temperature given by eq.~\eqref{hawkingtemp}.

As can be seen in figure~\ref{phasetrans1},
there is indeed a stability change, and the transition occurs at $L\sim 1.28$, which is the first marginally 
stable mode. In figure~\ref{fig_freq}, we show the real and imaginary parts of the quasi-normal frequency. 
Both parts change signal at the same value of $L$. 

\begin{figure}[htbp]
\includegraphics[height=7.0cm, width=4.5cm,angle=-90]{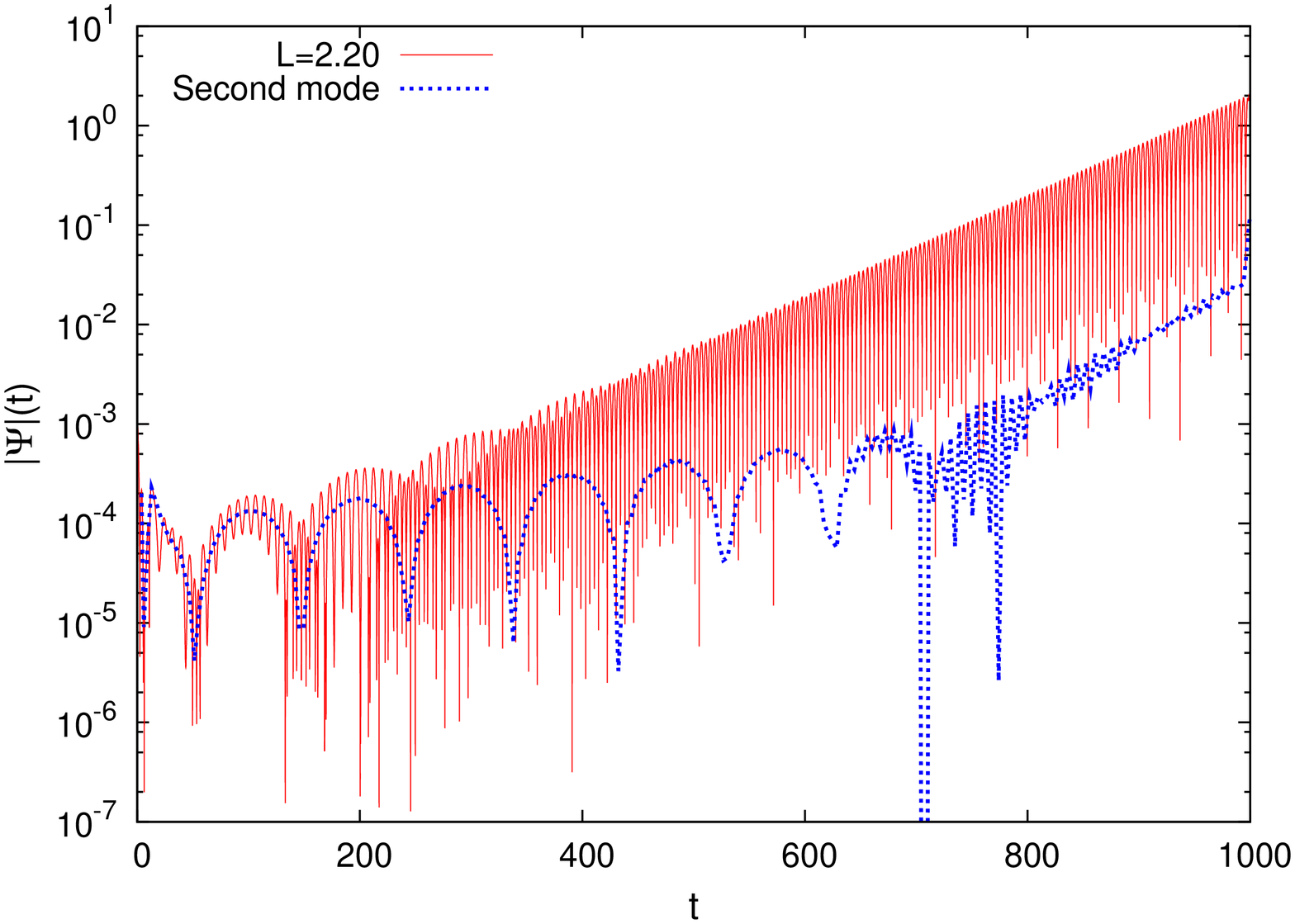}
\includegraphics[height=7.0cm, width=4.5cm, angle=-90]{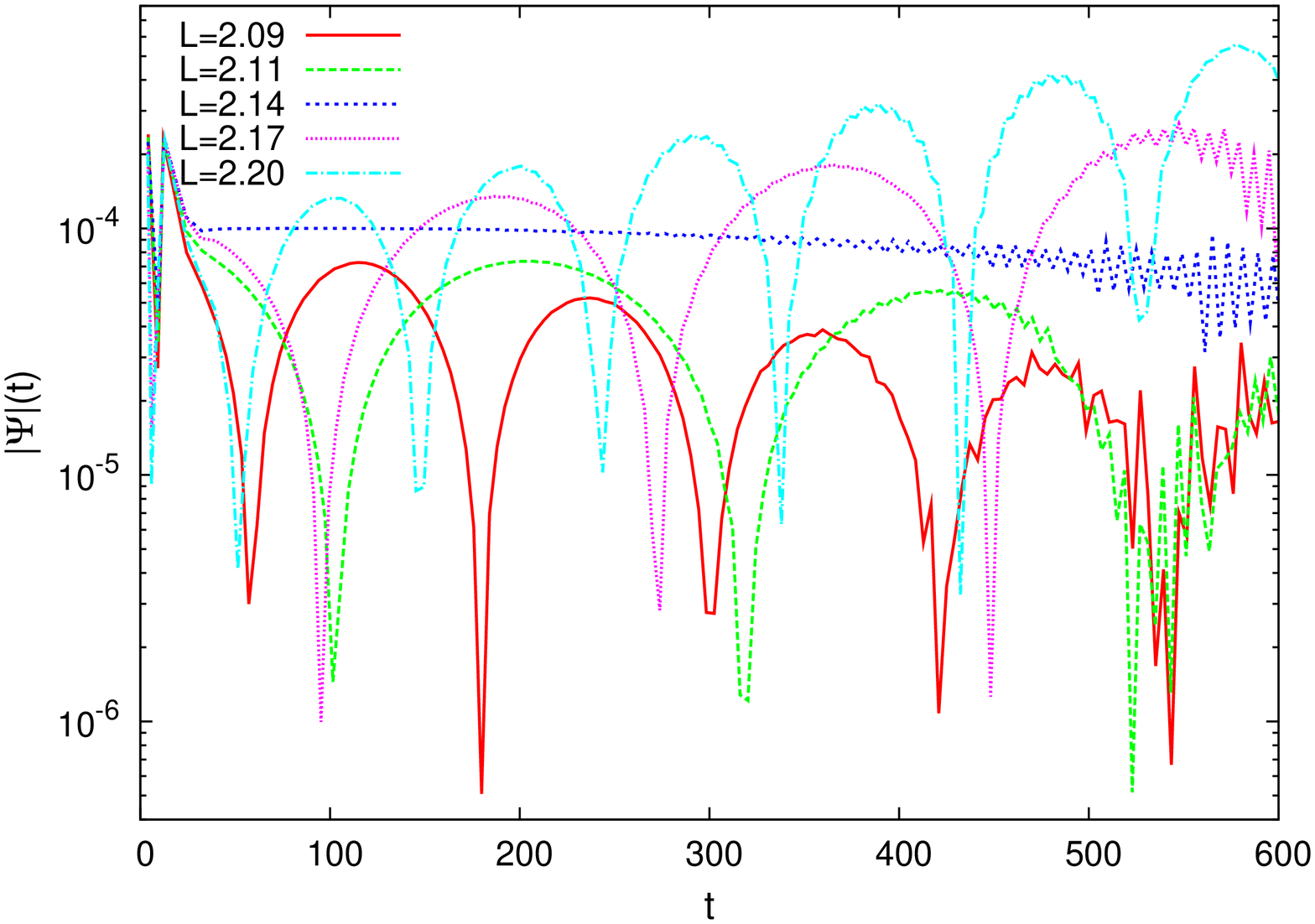}
\caption{{\bf Extracted secondary quasi-normal mode (left) and secondary quasi-normal modes for values of $L$ near the second marginally stable mode (right).}}
\label{phasetrans2}
\end{figure}

\begin{figure}[htbp]
\includegraphics[height=7.0cm, width=4.5cm,angle=-90]{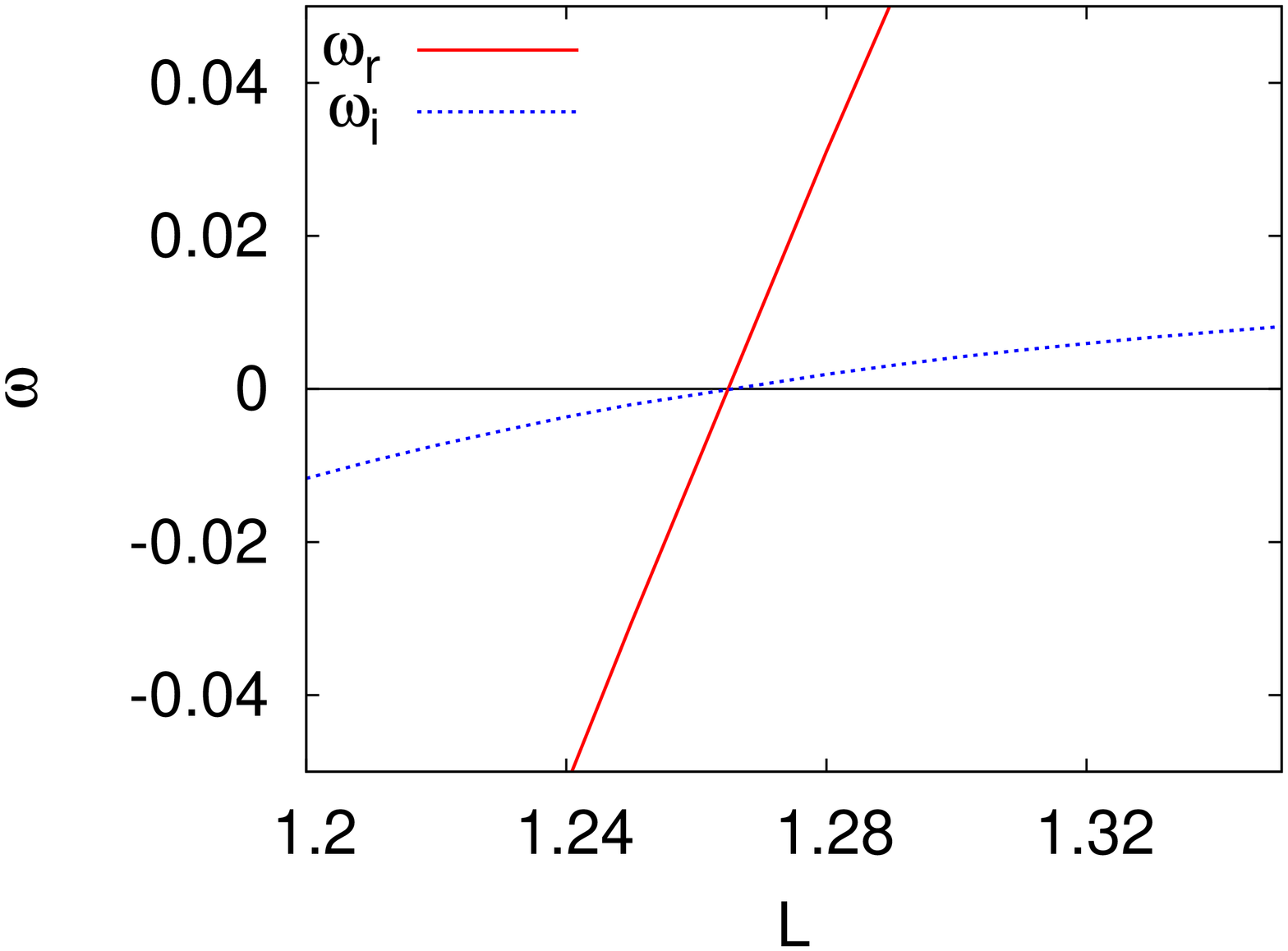}
\includegraphics[height=7.0cm, width=4.5cm,angle=-90]{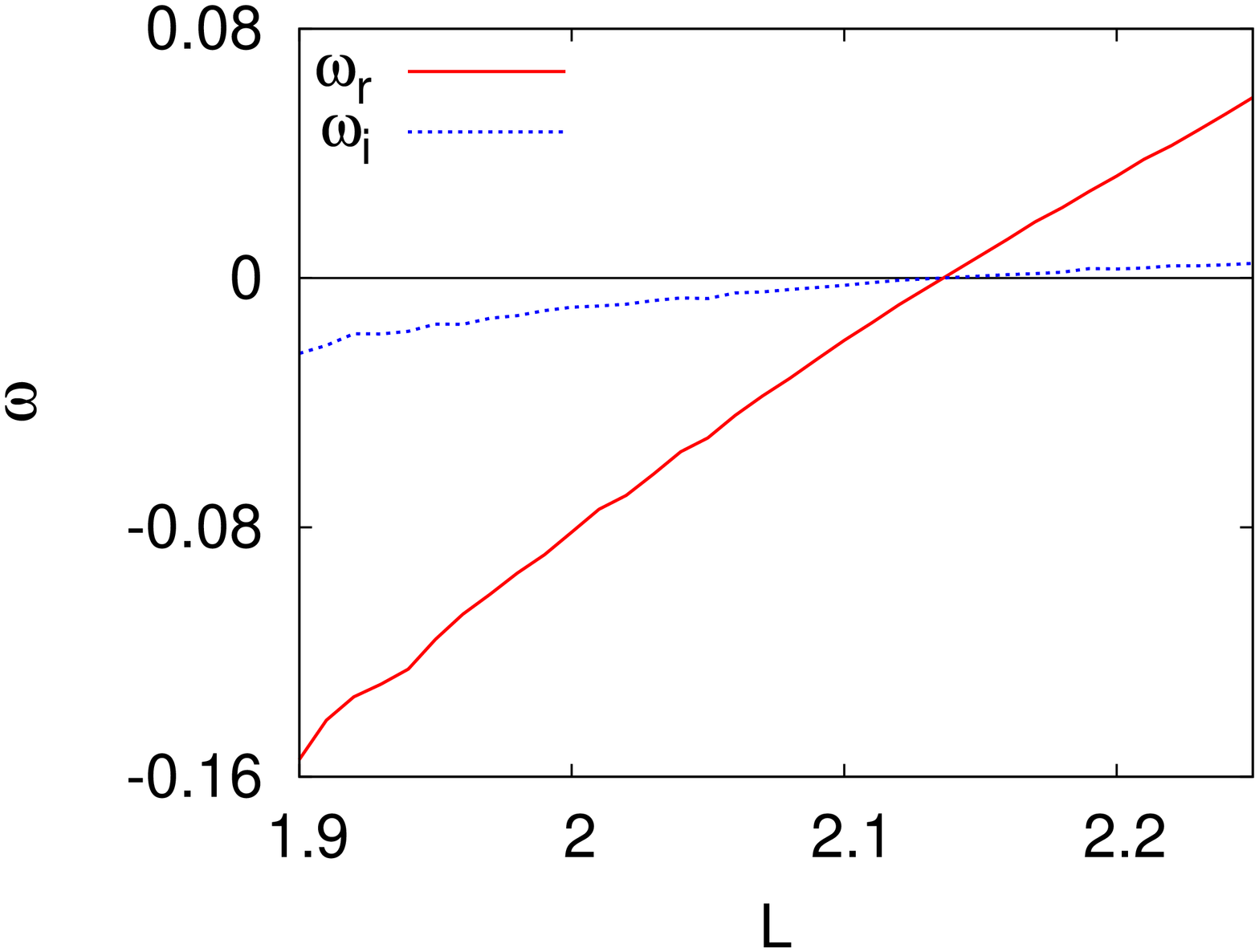}
\caption{{\bf Quasi-normal frequencies of fundamental mode (left) and of the extracted secondary mode (right).}}
\label{fig_freq}
\end{figure}

For values of $L$ close to the second marginally stable mode, $\psi(r,t)$ behaves as a
superposition of two oscilations with distinct frequencies. Figure~\ref{phasetrans2}
shows $\psi(r,t)$ and the extracted second mode. The second mode exhibits a transition at $L\sim 2.15$, which is
the second marginally stable mode. The frequencies are shown in figure~\ref{fig_freq}.
Since there is a third marginally stable mode, we assume that there is a third 
quasi-normal mode, but our analysis could not see it, as it is small compared to the other quasi-normal modes.

We can conclude now that the existence of more than one marginally stable mode means that there are more than one 
quasi-normal mode that changes stability at a particular marginally stable mode. For the system to be stable, all
quasi-normal modes must be stable, and when one single quasi-normal mode changes stability, the whole system is unstable.
So there is only one phase transition, which happens at the marginally stable mode labelled as fundamental.
This work has been published in \cite{phasetrans}.


\acknowledgments
This work has been supported by FAPESP and CNPq, Brazil.



\begin{thebibliography}{99}

\bibitem{maldacena1} J. Maldacena, 
{\it The Large N Limit of Superconformal Field Theories and Supergravity},
Adv. Theor. Math. Phys. {\bf 2} (1998) 231 [{\tt hep-th/9711200}].

\bibitem{witten} E.~Witten, 
{\it Anti De Sitter Space And Holography},
Adv. Theor. Math. Phys. {\bf 2} (1998) 253 [{\tt hep-th/9802150}].

\bibitem{1H} G.~T.~Horowitz, 
{\it Introduction to Holographic Superconductors}
[{\tt hep-th/1002.1722}].

\bibitem{gubser} S.~S.~Gubser, 
{\it Breaking an Abelian gauge symmetry near a black hole horizon}
Phys. Rev. D {\bf 78} (2008) 065034 [{\tt hep-th/0801.2977}].

\bibitem{phasetrans} E.~Abdalla, C.~E.~Pellicer, J.~Oliveira and A.~B.~Pavan, 
{\it Phase transitions and regions of stability in Reissner-Nordstr\"om holographic superconductors},
Phys. Rev. D {\bf 82} (2010) 124033 [{\tt hep-th/1010.2806}].

\bibitem{3Hs} S.~A.~Hartnoll, C.~P.~Herzog, and G.~T.~Horowitz, 
{\it Building an AdS/CFT superconductor},
Phys. Rev. Lett. {\bf 101} (2008) 031601 [{\tt hep-th/0803.3295}].

\bibitem{thooft} G.~'t Hooft, 
{\it Dimensional Reduction in Quantum Gravity},
[{\tt gr-qc/9310026}].

\bibitem{susskind} L.~Susskind, 
{\it The World as a Hologram},
J. Math. Phys. {\bf 36} (1995) 6377 [{\tt hep-th/9409089}].

\bibitem{nastase} H.~Nastase, 
{\it Introduction to AdS-CFT},
[{\tt hep-th/0712.0689}].

\bibitem{mcgreevy} J.~McGreevy, 
{\it Holographic duality with a view toward many-body physics},
[{\tt hep-th/0909.0518}].

\bibitem{kokkotas}
K.~D.~Kokkotas and B.~Schmidt, 
{\it Quasi-Normal Modes of Stars and Black Holes}, 
Living Rev. Relativity {\bf 2} (1999).

\bibitem{nollert} H.-P.~Nollert, 
{\it Quasinormal modes: the characteristic `sound' of black holes and neutron stars},
Class. Quantum Grav. {\bf 16} (1999) R159.

\bibitem{wald} R.~M.~Wald, 
{\it Note on the stability of the Schwarzschild metric}, 
J. Math. Phys. {\bf 20} (1979) 1056.

\bibitem{2Hs}
G.~T.~Horowitz and V.~E.~Hubeny, 
{\it Quasinormal modes of AdS black holes and the approach to thermal equilibrium}, 
Phys. Rev. D {\bf 62} (2000) 024027. [{\tt hep-th/9909056}].

\bibitem{franceses}
A. Bachelot e A. Motet-Bachelot, 
{\it Les r\'esonances d'un trou noir de Schwarzschild.}, 
Ann. Inst. Henri Poincar\'e {\bf 59} (1993) 3-68.

\bibitem{gibbons} G. W. Gibbons and P. K. Townsend, 
{\it Vacuum interpolation in supergravity via super p-branes}, 
Phys. Rev. Lett. {\bf 71} (1993) 3754 [{\tt hep-th/9307049}].

\bibitem{pbranas} E.~Abdalla, O.~P.~F.~Piedra and J.~de Oliveira, C. Molina
  {\it Perturbations of black p-branes},
  Phys.\ Rev.\ D {\bf 81} (2010) 064001 [{\tt hep-th/0810.5489}].

\bibitem{kodama} H. Kodama, A. Ishibashi, 
{\it A master equation for gravitational perturbations of maximally symmetric black holes in higher dimensions},
Prog. Theor. Phys. {\bf 110} (2003) 701 [{\tt hep-th/0305147}].

\bibitem{agmoo} O.~Aharony, S.~S.~Gubser, J.~Maldacena, H.~Ooguri and Y.~Oz, 
{\it Large N Field Theories, String Theory and Gravity},
Phys.Rept. {\bf 323} (2000) 183-386 [{\tt hep-th/9905111}].


\end{thebibliography}
\end{document}